\documentclass[pra,twocolumn,superscriptaddress,showpacs,floatfix]{revtex4-1}
\usepackage{graphics}
\usepackage{amsmath}
\usepackage{braket}
\usepackage{graphicx}
\usepackage{float}
\usepackage{color}
\usepackage{rotating}
\usepackage{tablefootnote}
\usepackage{threeparttable}

\newcommand{\var} [1] [r]{
\mathbf{#1}
}

\usepackage{tikz}
\usetikzlibrary{shapes.geometric, arrows}

\tikzstyle{process} = [rectangle, minimum width=3cm, minimum height=1cm, text centered, text width=5cm,draw=black] 
\tikzstyle{decision} = [diamond, minimum width=5cm, minimum height=1cm, text centered, text width=3cm, draw=black]
\tikzstyle{arrow} = [thick,->,>=stealth]

\begin{document}
\title{Performance of the Constrained Minimization of the Total Energy in Density Functional Approximations: the Electron Repulsion Density and Potential}
\author{Tom Pitts}
\affiliation{Department of Physics, Durham University, South Road, Durham, DH1 3LE, United Kingdom}
\author{Nikitas I. Gidopoulos}
\affiliation{Department of Physics, Durham University, South Road, Durham, DH1 3LE, United Kingdom}
\author{Nektarios N. Lathiotakis}
\affiliation{Theoretical and Physical Chemistry Institute, National Hellenic Research Foundation, Vass.~Constantinou 48, GR-116 35, Athens, Greece}

\date{\today}
%
\begin{abstract}
In the constrained minimization method of Gidopoulos and Lathiotakis 
(J. Chem. Phys. {\bf 136}, 224109), the Hartree exchange and 
correlation Kohn-Sham potential of a finite $N$-electron system is replaced by 
the electrostatic potential of an effective charge density that is everywhere 
positive and integrates to a charge of $N-1$ electrons. The optimal effective 
charge density (electron repulsion density, $\rho_{\rm rep}$) 
and the corresponding optimal effective potential (electron repulsion potential 
$v_{\rm rep}$) are obtained by 
minimizing the electronic total energy in any density functional approximation.
The two constraints are sufficient to remove the self-interaction errors from 
$v_{\rm rep}$, correcting its asymptotic behavior at large distances from the 
system.  
In the present work, we describe, in complete detail, the constrained minimization method,
including recent refinements.
We also assess its performance in removing the self-interaction errors 
for three popular density functional approximations, namely LDA, PBE and B3LYP, 
by comparing the obtained ionization energies to their experimental values for an 
extended set of molecules. 
We show that the results of the constrained minimizations are almost independent 
of the specific approximation with average percentage errors 15\%, 14\%, 13\% for 
the above DFAs respectively. These errors are substantially smaller than the 
corresponding errors of the plain (unconstrained) Kohn-Sham calculations at 
38\%, 39\% and 27\% respectively. Finally, we showed that this method correctly predicts
negative values for the HOMO energies of several anions.
\end{abstract}
\maketitle
%
%
%
\section{Introduction}
\label{intro}

It is well known that approximations in density functional theory (DFT) suffer from self-interaction (SI) errors 
\cite{pzsic}. In the total energy, SIs arise in the Hartree (or Coulomb) 
term that represents the electrostatic Coulomb repulsion energy of the electronic charge density $\rho$ with itself. 
In theories that employ a non-interacting $N$-particle (Slater determinant) state to represent the interacting system, 
(like Hartree-Fock, Kohn-Sham-DFT), the charge density $\rho$ is the sum of the single-particle densities of the 
orbitals that form the Slater determinant. 


In Hartree-Fock theory, this self repulsion is cancelled exactly for the occupied orbitals~\cite{hhfj}, by the Fock exchange term. 
In KS DFT, the same happens with the exact exchange energy functional, $E_x [\rho]$, 
which is also based on the Fock exchange energy expression in terms of the KS orbitals. 
However, for approximate exchange energy functionals the cancellation of the SIs is not complete.

Self interactions have a large impact on the accuracy of many properties predicted by density functional approximations. 
These errors include: artificial stabilization of delocalized states \cite{lundberg2005quantifying},  underestimating electron 
affinities \cite{rosch1997comment} and the underestimation of ionization energies and band 
gaps\cite{toher2005self,goedecker1997critical,perdew1983physical}. 

The SI error is readily observed in the asymptotic 
behavior of the Kohn Sham (KS) potential \cite{vonbarth}. 
For an $N$-electron system, in a theory without SIs, 
the electron-electron part of the KS potential should decay at a large distance $r$ away from the system 
as $(N-1) / r$, corresponding to the electrostatic potential of a charge of $N-1$ electrons. 
In DFT, the electron-electron interaction is given by the sum 
of the Hartree potential $v_H ({\bf r})$ and the exchange and correlation potential $v_{xc} ({\bf r})$.
The asymptotic decay of the Hartree potential is $N / r$ and 
the exchange and correlation potential decays as $-1 / r $ 
in a SI free approximation. 
Hence, SIs are evident when $v_{xc} ({\bf r})$ does not decay as $- 1 / r$ 
and in many popular density functional approximations (DFAs) $v_{xc} ({\bf r})$ is found 
to decay exponentially fast. 
The result is that in these approximations, the Hartree, exchange and 
correlation (Hxc) part of the KS potential, $v_{\rm Hxc} ({\bf r})$,  
decays as $N / r $. 
This asymptotic behavior of $v_{\rm Hxc} ({\bf r})$ 
reveals that an electron of the system interacts with the 
charge density of all the electrons in the system including itself. 

To expand on this point, 
Poisson's law can be used \cite{Gor1999,gidopoulos2012constraining} to define the 
charge density (denoted by $\rho_{\rm Hxc}$), whose electrostatic potential is $v_{\rm Hxc} ({\bf r})$:
\begin{equation}
\nabla^2 v_{\rm Hxc} ({\bf r}) = - 4 \pi \rho_{\rm Hxc} ({\bf r}) , 
\  v_{\rm Hxc} ({\bf r}) = \int {d{\bf r}' \,  \rho_{\rm Hxc} ({\bf r}' ) \over |{\bf r} - {\bf r} ' | } \,.
\end{equation}
Then, the presence of SIs in the approximate KS potential of a finite 
system can be quantified in terms of the integrated charge of 
$\rho_{\rm Hxc} ({\bf r})$ \cite{gidopoulos2012constraining,GIDOPOULOS2015129}. If $\int d{\bf r} \rho_{\rm Hxc} ({\bf r}) = N-1$, 
the approximate KS potential is free from SIs, while 
if $\int d{\bf r} \rho_{\rm Hxc} ({\bf r}) = N$, then there are full SIs in the approximation.


There have been several attempts to correct for SI effects 
\cite{pzsic,lundberg2005quantifying,toher2005self,goedecker1997critical,LB94,ADSIC,gidopoulos2012constraining,tsuneda2014self,pederson2014communication,clda_review,clark_gidopoulos_sic}. 
The best known is the method proposed by Perdew and Zunger in 1980 
(PZ-SIC) \cite{pzsic}, in which the 
SI error for each orbital is subtracted from the total energy, 
yielding a SI corrected total energy expression. 
A drawback of the PZ-SIC method is that its SI correction term breaks 
the invariance of the total energy w.r.t unitary 
transformations of the occupied orbitals, an issue that was addressed 
recently by Perdew and co-workers~\cite{pederson2014communication}. 
In addition, there is a number of independent 
SI corrections that keep unitary invariance of 
the occupied orbitals, for a list see Ref.~\cite{kummel_perdew}. 

A method for correcting for SI effects in the KS potential (but without correcting the energy) 
was proposed by Gidopoulos and Lathiotakis~\cite{gidopoulos2012constraining,GIDOPOULOS2015129}. 
In place of $v_{\rm Hxc}({\bf r})$, it employs a different effective local potential to represent the electronic repulsion, 
denoted by $v_{\rm rep}({\bf r})$. 
The latter is variationally optimized, under two constraints, which affect the effective potential everywhere, forcing it to exhibit the correct asymptotic tail $(N - 1) / r$ at large distances from the system.
The novelty of this proposition is the constrained variational optimization of the effective potential for DFAs (like LDA, GGA or hybrid), for which the usual KS scheme would normally be employed to obtain the minimum of the total energy in an unconstrained manner. By employing these constraints in the optimization process, it becomes possible to incorporate in the resulting effective potential properties of the exact KS potential that these approximations would otherwise violate.

Since the potential is obtained variationally, the proposition of 
Ref.~\cite{gidopoulos2012constraining} 
is similar to the OEP method. However, until 
Ref.~\cite{gidopoulos2012constraining}, the OEP method had been employed 
for the 
minimization of implicit density functional (orbital functionals), like exact 
exchange, and not for the more common DFAs that are explicit functionals of 
the density, as LDA or GGA.

In Ref.~\cite{gidopoulos2012constraining}, the method was shown to correct 
the asymptotics of the effective KS potential and gave improved results for 
the ionization potentials (IPs), compared with experiment. These improvements 
were demonstrated for the local density approximation (LDA) and for a small 
set of atoms and molecules. 
In addition, in order to capture both static correlation effects (using 
fractional occupations) as well as one-electron properties (from the KS 
spectrum), the constrained minimization technique 
of \cite{gidopoulos2012constraining} was employed in the indirect 
minimization of the total energy, expressed as a functional of the one-body, 
reduced, density 
matrix \cite{localrdmft,localrdmftappl,GIDOPOULOS2015129,lrdmftorbs}.

In the present work we describe in complete detail the 
constrained minimization method including recent refinements.
We also validate our method and demonstrate its applicability with two
 additional popular DFAs, the functional by Perdew, Burke, Ernzerhof 
(PBE) \cite{perdew1996generalized} and the B3LYP hybrid 
functional~\cite{becke1993density,lee1988development}.
Thus, we obtain similarly improved results for the IPs of an 
extended set of molecules, with the three DFAs: LDA, PBE and B3LYP. 
The IP is found as the negative of the energy eigenvalue 
of the highest occupied molecular orbital 
(HOMO)\cite{zhan2003ionization}, 
a quantity that is sensitive to the effects of SIs. 
These calculations are carried out for both the unconstrained and 
constrained methods and are compared to experimental 
results for the IP. 

\section{Method}
\label{sec:method}

In Ref.~\cite{gidopoulos2012constraining}, the Hartree, 
exchange and correlation potential $v_{\rm Hxc}$ in the KS equations is replaced by 
an effective potential $v_{\rm rep}$ that simulates the repulsion between the electrons (similarly to $v_{\rm Hxc}$). 
The single-particle (KS) equations take the form: 
\begin{equation} 
\left[ -\frac{\nabla^2}{2} + v_{en}(\var) + v_{\rm rep}(\var) \right] \phi_i(\var) =\epsilon_i \phi_i(\var) , 
\label{modified_KS}
\end{equation}
where $v_{en}$ is the attractive electron-nuclear potential. 
The density of the $N$ lowest orbitals of \eqref{modified_KS} is
\begin{equation} \label{den}
\rho (\var) = \sum_{i = 1}^N | \phi_{ i} (\var) |^2\,. 
\end{equation}
The effective potential $v_{\rm rep}$ is then represented as the electrostatic potential of an effective charge density $\rho_{\rm rep}$ giving rise 
to electron repulsion,
\begin{equation}
v_{\rm rep}(\var) = \int  \frac{ d\var[r'] \, \rho_{\rm rep}(\var[r'])}{|\var-\var[r']|} . \label{rep_pot_eq}
\end{equation}
In order to correct SIs, 
the following conditions are imposed on the effective electron repulsion density $\rho_{\rm rep}$: 
\begin{equation}
\int \rho_{\rm rep}(\var) \, d\var \  = N-1, \label{constraint1}
\end{equation}
\begin{equation}
\rho_{\rm rep}(\var) \geq 0.  \label{constraint2}
\end{equation} 
The normalization constraint in \eqref{constraint1} is a necessary condition satisfied by the
exact KS potential. When it is satisfied the potential has the correct asymptotic behavior.
This condition has been considered previously by G\"orling\cite{Gor1999}
in the framework of exact exchange OEP. In that case, it was employed 
to correct inaccuracies related to the finite basis expansion of the orbitals
 and of the potential, since the exact exchange potential is correct in the asymptotic region, but only for a complete basis. 

The constraint \eqref{constraint1} on its own is not sufficient to yield physical potentials:
in the minimization of the DFA total energy, 
it would be energetically favorable to yield the charge density corresponding to Hxc potential of the DFA 
($v_{\rm Hxc}^{\rm DFA}$, which decays exponentially fast), 
combined with an opposite charge of $-1$ spread out at a large distance away from the electronic system.
Introducing the additional positivity constraint \eqref{constraint2} ensures that the mathematical problem of 
determining $\rho_{\rm rep}$ becomes well posed. 
The two constraints, \eqref{constraint1}, \eqref{constraint2}, affect the electron repulsion density over all 
space and not just in the asymptotic region away from the molecule; hence these constraints do not merely correct the asymptotic tail of the 
electron repulsion potential.   
 
It should be noted that the potential $v_{\rm rep}$, which plays the role of $v_{\rm Hxc}$ in the KS equations, 
is not defined as the functional derivative of the approximate Hxc energy w.r.t.~the 
density.


To proceed, we seek the effective potential $v_{\rm rep}$ in Eq.~\eqref{modified_KS}, 
whose orbitals $\phi_i$ give the density 
$\rho$ (Eq.~\eqref{den}) that minimizes the DFA total energy,
\begin{equation} \label{te}
E_{v_{en}}^{\rm DFA} [\rho] = T_s [\rho] + \int d \var \, v_{en} (\var) \, \rho(\var) + E_{\rm Hxc}^{\rm DFA} [\rho]\,, 
\end{equation}
where $E_{\rm Hxc}^{\rm DFA} [\rho]$ is the Hxc energy functional of the density in the DFA.
Since, the density in \eqref{te} depends on the ($N$-lowest) orbitals of $v_{\rm rep}$, 
the total energy becomes a functional of $v_{\rm rep}$.
The functional derivative of the total energy w.r.t. the potential is:
\begin{equation} \label{fd}
{\delta E_{v_{en}}^{\rm DFA} [ v_{\rm rep} ]\over \delta v_{\rm rep} (\var) } =  
\int  d \var ' \, \chi(\var, \var[r']) \, \Big[ v_{\rm Hxc}^{\rm DFA} [\rho] (\var ' ) - v_{\rm rep} (\var ') \Big]\,, 
\end{equation}
where $\chi (\var , \var ' )$ is the density response function,
\begin{equation}
\chi (\var , \var ' ) = 
\sum_{i}^{\rm occ} \sum_{a}^{\rm unocc} 
\frac{ \phi_i (\var) \, \phi_a^* (\var) \, \phi_i^* (\var ' ) \, \phi_a (\var ' )   }{\epsilon_{ i } - \epsilon_{ a }} 
+ {\rm c.c.} \, , 
\label{rf}
\end{equation}
$\phi_{ k}, \epsilon_{ k}$ are the KS orbitals and energies in \eqref{modified_KS}
and 
\begin{equation}
v_{\rm Hxc}^{\rm DFA} [\rho ] (\var) = \left. \frac{\delta E_{\rm Hxc}^{\rm DFA} [\rho]}{\delta \rho(\var)}\right|_{\rho} 
\end{equation}
is the Hartree, 
exchange and correlation potential of the DFA, evaluated at $\rho$. 

Since  $\chi$ does not have singular (or null) eigenfunctions apart from the constant function~\cite{talman_bartlett}, 
the effective potential $v_{\rm rep} (\var)$ for which the functional derivative \eqref{fd} vanishes 
is $v_{\rm rep} = v_{\rm Hxc}^{\rm DFA} [\rho ]$, modulo a constant function.
It is reassuring that before imposing the two constraints \eqref{rep_pot_eq}-\eqref{constraint2}, 
the variationally optimal potential from the minimization of the total energy turns out to be the Hxc potential of the 
DFA, as expected. 
It is worth noting that up to this point, our total energy minimization follows the optimized effective potential (OEP) method, even when we employ a benign approximation (such as LDA/PBE) for the XC energy functional. 
We now proceed to enforce the two constraints on the effective potential, which is where we deviate from the OEP methodology.

Compared with Ref.~\cite{gidopoulos2012constraining}, in the present work, we have modified slightly  
the way we enforce the positivity constraint \eqref{constraint2}. 
In this work, to implement the two constraints \eqref{rep_pot_eq}-\eqref{constraint2}, we employ a Lagrange multiplier $\lambda$ to
satisfy \eqref{constraint1}, and a penalty term that increases the energy of the objective function in all points $\var$ where the 
effective charge density $\rho_{\rm rep}(\var)$ is negative. The Lagrange multiplier $\lambda$ and the penalty coefficient $\Lambda$ have units of energy.
Since the effective potential $v_{\rm rep}$ depends on the effective density $\rho_{\rm rep}$, the energy becomes a functional of $\rho_{\rm rep}$ 
and the objective quantity to be minimized becomes:
\begin{multline}
G_{v_{en}} [ \rho_{\rm rep}]  =
E_{v_{en}}^{\rm DFA} [ \rho_{\rm rep}]  - \lambda \left[\int d\var \, \rho_{\rm rep}(\var) - (N-1) \right] \\
+ \Lambda\left[\int d\var \ |\rho_{\rm rep}(\var)|  - (N-1) \right]\,. \label{obj}
\end{multline}
At the minimum of $G_{v_{en}}$, the derivative must vanish:
\begin{multline}
 \int {d \var  \over | \var - \var[x] | } \int d \var ' \chi(\var,\var[r'])  
  \big[ v_{\rm Hxc}^{\rm DFA}(\var') - v_{\rm rep}(\var') \big]  
\\
 - \lambda + \Lambda \, \mathrm{sgn} [\rho_{\rm rep}(\var[x]) ] = 0 \,,
\end{multline}
where $\mathrm{sgn}[x]$ is the signum function.

Introducing
\begin{equation}
\tilde b({\bf x} ) = \int  {d \var  \over | \var - {\bf x}| } \int d \var '   \, \chi(\var,\var[r']) \,  
v_{\rm Hxc}^{\rm DFA}(\var ' ) , 
\label{tb}
\end{equation}
and
\begin{equation}
\tilde \chi ({\bf x} , {\bf y}) =  \int \! \! \int {d \var \, d \var ' \, \chi(\var,\var[r'])  \over | \var - \var[x] | \,  | \var ' - {\bf y}|   } \,, 
\label{trf}
\end{equation}
the equation determining the effective density $\rho_{\rm rep}(\var[x])$ becomes: 
\begin{equation} \label{lin}
  \int d {\bf y} \, \tilde \chi ({\bf x} , {\bf y}) \,  \rho_{\rm rep} ({\bf y})  
= \tilde b ({\bf x}) 
 - \lambda + \Lambda \, \mathrm{sgn} \left[\rho_{\rm rep}(\var[x]) \right] .
\end{equation}
We expand $\rho_{\rm rep}(\var)$ in the auxiliary basis $\{ \chi_n (\var)\}$,  
\begin{equation}
\rho_{\rm rep}(\var) = \sum_n \nu_n \, \chi_n(\var) \label{expansion}\,,
\end{equation}
and the optimization w.r.t. $\rho_{\rm rep} (\var)$ transform to the search for
the optimal expansion coefficients $\nu_n$. 
Substituting the expansion \eqref{expansion} into Eq.~\eqref{lin}, multiplying by $\chi_k(\var[x])$ and integrating over $\var[x]$, we have: \begin{multline} \label{leq}
 \sum_n \nu_n \iint  d\var[y] \, d\var[x] \, \chi_k (\var[x])  \tilde{\chi}(\var[x], \var[y]) \chi_n(\var[y])  = \\ 
 \int d\var[x] \,  \tilde{b}(\var[x]) \chi_k(\var[x])  \ - \lambda  \int d\var[x] \,  \chi_k(\var[x]) \\
 + \Lambda \int d\var[x] \, \chi_k(\var[x]) \, \mathrm{sgn} [\rho_{\rm rep}(\var[x]) ] \,.
\end{multline}
We define:
\begin{eqnarray}
A_{k n} &=& \iint d{\bf x} \, d{\bf y} \, \chi_k ({\bf x}) \, \tilde \chi ({\bf x} , {\bf y} ) \, \chi_n ({\bf y}) 
\label{mA} \\
b_k &=& \int d {\bf r} \, \tilde b ({\bf r}) \, \chi_k( \var ) \label{bk} \\
X_k & = & \int d \var \, \chi_k (\var) \label{Xk} \\
\bar X_k &=&  \int d\var \, \chi_k (\var ) \, \mathrm{sgn} [\rho_{\rm rep}(\var ) ]\,, 
\label{bXk}
\end{eqnarray}
and Eq.~\eqref{leq} becomes:
\begin{equation}
\sum_n A_{ k n} \, \nu_n = b_k - \lambda \, X_k + \Lambda \, \bar X_k\,.
\end{equation}
The solution is obtained by inverting the matrix $A_{k n}$:
\begin{equation} \label{leq1}
\nu_m = \sum_k A_{ m k}^{-1} \, b_k - \lambda \, \sum_k A_{ m k}^{-1} \, X_k + \Lambda \, \sum_k A_{ m k}^{-1} \, \bar X_k \,.
\end{equation}
From Eqs.~\eqref{constraint1}, \eqref{Xk} we have $\sum_m X_m \, \nu_m = N-1$. Then, we obtain for the Lagrange multiplier $\lambda$:
\begin{equation} \label{leq2}
 \lambda  = {\sum_{k , m} X_m \,  A_{ m k}^{-1} \,  \big[b_k  + 
\Lambda \, \bar X_k \big] - (N-1) \over \sum_{l , n} X_n \, A_{ n l}^{-1} \, X_l }\,.
\end{equation}
Eqs.~\eqref{leq1}, \eqref{leq2} determine the expansion coefficients $\nu_n$ of the effective charge density $\rho_{\rm rep}$. 

With a finite orbital basis, the matrix $A_{kn}$ has vanishingly small eigenvalues requiring a singular value decomposition 
(SVD) to remove the projections to the (almost) null eigenvalues from the matrix. 
The choice of cutoff point for the nonzero eigenvalues is often ambiguous. 
Including too many small, but non-zero, eigenvalues, leads to a slower and 
probably not convergent calculation, while omitting them 
might result in pure representation of the effective density. Both 
cases may result in small differences in the calculated HOMO energy. 
We found that a cutoff of $\sim 10^{-5}$ was a good choice for most molecules. 
For a better way to determine the cutoff point for the singular eigenvalues, see Ref.~\cite{gl_nonanalyticity}. \\


\begin{figure}
\begin{center}
\begin{tikzpicture}[node distance=1.5cm, every node/.style={scale=0.8}]
\node (start) [process] {Start with initial guess for the KS orbitals};
\node (n1) [process, below of=start] {From the KS orbitals find $\tilde \chi$, $A_{k n}$, $b_k$, $ X_k$, $\bar X_k$ };
\node (n2) [process, below of=n1] {Keeping the orbitals fixed calculate $\nu_m$ via \eqref{leq1}};
\node (n3) [process, below of=n2] {Use  $\nu_m$ to find $\rho_{\rm rep}$ from \eqref{expansion} };
\node (d1) [decision, below of=n3, yshift=-1.9cm] {Is the negative component of $\rho_{\rm rep}$ sufficiently small? };
\node (n4) [process, below of=d1, yshift=-1.7cm] {Recalculate $v_{\rm rep}$ from $\rho_{\rm rep}$ };
\node (n5) [process, below of=n4] {Update KS orbitals and calculate total energy};
\node (d2) [decision, below of=n5, yshift=-1.7cm] {Has the energy converged sufficiently};
\node (end) [process, below of=d2, yshift=-1.7cm] {Calculate final properties};

\node (fic_d1) [right of=d1, xshift=1.5cm] {};
\node (fic_d2) [left of=d2, xshift=-1.5cm] {};

\draw [arrow] (start) -- (n1);
\draw [arrow] (n1) -- (n2);
\draw [arrow] (n2) -- (n3);
\draw [arrow] (n3) -- (d1);
\draw [arrow] (d1) -- node [right] {yes} (n4);
\draw [arrow] (n4) -- (n5);
\draw [arrow] (n5) -- (d2);
\draw [arrow] (d2) -- node [right]{yes} (end);

\draw [arrow] (d1) -- node [below] {no} (fic_d1.center) |- (n2);
\draw [arrow] (d2) -- node [below]{no} (fic_d2.center) |- (n1);

\end{tikzpicture}
\end{center}
\caption{\label{fig:flow}
A flow diagram showing the procedure for a constrained calculation. } 
\end{figure}
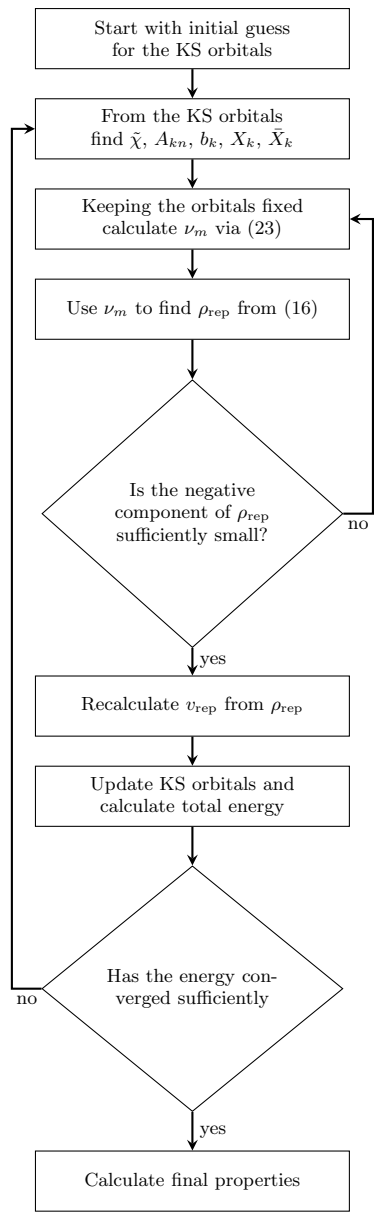



In our iterative procedure, shown in Fig.~\ref{fig:flow}, we do not need an initial guess for $\rho_{\rm rep}$. Instead we start from an initial guess for the KS orbitals, e.g. the LDA orbitals. 
From these orbitals we calculate $\tilde{\chi}$, $A_{kn}$, $b_k$, $X_k$, ${\bar X}_k$ and the initial $\rho_{\rm rep}$. 
From Eq.~\eqref{rep_pot_eq}, we obtain the effective potential
$v_{\rm rep}$ and solve the KS equations \eqref{modified_KS}.
In the inner loop, with these KS orbitals and eigenvalues we find the 
response functions $\chi$ of Eq.~\eqref{rf} and $\tilde \chi$ of Eq.~\eqref{trf} and the matrix $A_{k n}$ in Eq.~\eqref{mA}.  

We also find the $N$-electron density $\rho$, the potential $v_{\rm Hxc}^{\rm DFA} [\rho] $,
the function $\tilde b$ of Eq.~\eqref{tb}, and its projections on the auxiliary basis functions \eqref{bk} and finally $\bar X_k$ of Eq.~\eqref{bXk}. 
The latter differ from $X_k$ of Eq.~\eqref{Xk} when the effective density changes sign.  
Using all these, we update the effective density (Eq.~\eqref{expansion}) by solving \eqref{leq1}. Still in the inner loop, 
keeping orbitals and eigenvalues fixed, we update iteratively $\bar X_k$  and the effective density (Eqs.~\eqref{expansion}, \eqref{leq1}), 
until the following measure of negativity of $\rho_{\rm rep}$
\begin{equation}
Q_{\rm neg} = \int d\var[r]\; \big[ \rho_{\rm rep}(\var[r]) - |\rho_{\rm rep}(\var[r])|\big]
\end{equation}
is sufficiently small. Practically, a criterion for positivity $Q_{\rm neg} < 10^{-6}$ was used.
In the inner loop we used a mixing scheme for $\rho_{\rm rep}$ and the efficiency/convergence were
controlled by the values of the penalty parameter, $\Lambda$, and the mixing parameter, $x_m$. 
Typically, for $\Lambda$, we used a value of the order of $\sim 100$~a.u. combined with a very 
small starting value for $x_m$ ($\sim 10^{-8}$) which was dynamically raised or lowered based on the change of $Q_{\rm neg}$ at each successive iteration.
This loop is the bottleneck of our method at the present stage.
An update of our method that enforces positivity in a direct and more 
efficient way is work in progress. When the positivity criterion is 
satisfied, we recalculate the effective potential $v_{\rm rep}$ of 
Eq.~\eqref{rep_pot_eq}, solve the KS equations and iterate the outer 
loop with updated orbitals.



\section{Results}

This method was implemented in the code HIPPO\cite{lathiotakis2008benchmark} 
using Gaussian basis sets to expand both the orbitals and the potentials; for the expansion of the orbitals we chose 
the cc-pVDZ basis sets as a good compromise between accuracy and speed for the calculations.
Pairing the orbital basis with the corresponding uncontracted for the auxiliary basis to expand 
$\rho_{\rm rep}$ was proven a successful combination for all tests we have performed.

To demonstrate the improvement of the constrained method vs the unconstrained approach, the highest occupied molecular 
orbital (HOMO) energies of large number of molecules were calculated and compared to experimental results for the 
IPs from the NIST computational chemistry comparison and benchmark database (CCCBDB)~\cite{johnson2005nist}. 
To show the applicability of our method to different approximations, three DFAs were investigated, LDA, PBE, 
and the hybrid functional B3LYP. These DFAs are among the most popular functionals for electronic structure calculations
and they all contain self-interaction effects, to some degree.

\begin{figure}
\begin{center}	
\includegraphics[ width=0.45\textwidth]{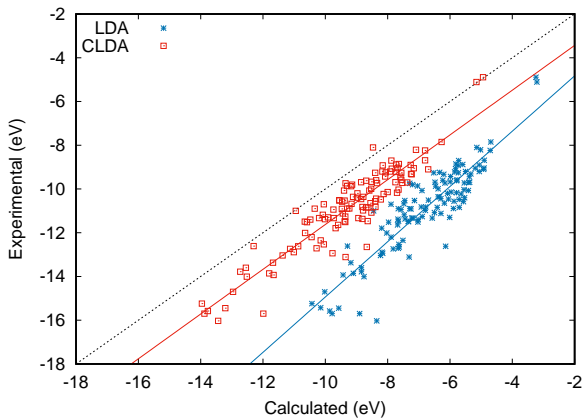} 
\caption{Calculated IPs using the LDA compared with experimental values. 
Blue stars show results from unconstrained minimization; red boxes show results of the constrained minimization. 
Red and blue lines are guides to the eye. The IP is found as the negative of the HOMO energy. 
The black dotted line corresponds to the ideal correlation between an exact calculation and experiment.} \label{LDA_graph}
\end{center}
\end{figure}

\begin{figure}
\begin{center}	
\includegraphics[width=0.45\textwidth]{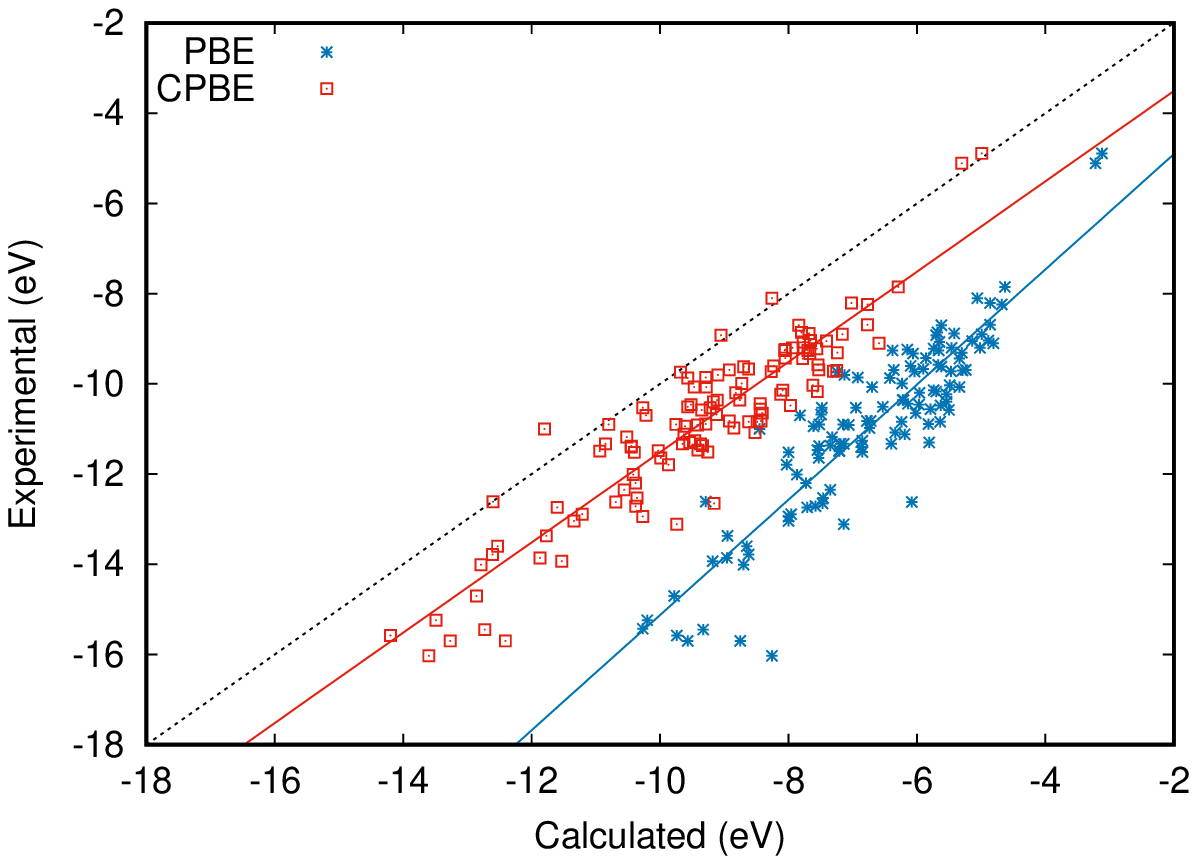} 
\caption{Calculated IPs using the PBE functional compared with experimental values. 
Blue stars show results from unconstrained minimization; red boxes show results of the constrained minimization. 
Red and blue lines are guides to the eye. The IP is found as the negative of the HOMO energy. 
The black dotted line corresponds to the ideal correlation between an exact calculation and experiment.} \label{PBE_graph}
\end{center}
\end{figure}

\begin{figure}
\begin{center}	
\includegraphics[width=0.5\textwidth]{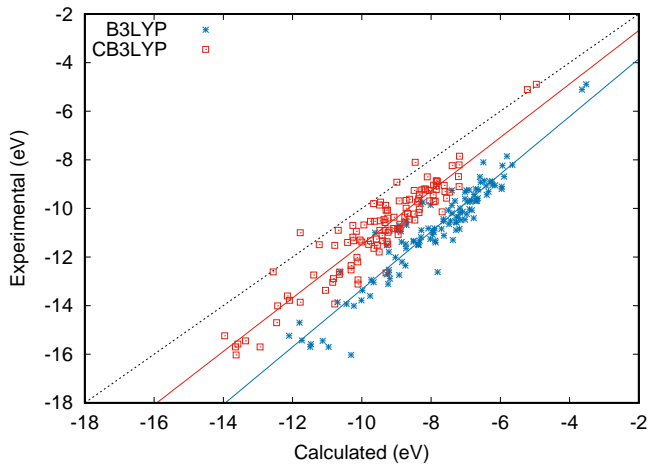} 
\caption{Calculated IPs using the B3LYP hybrid functional compared with experimental values. 
Blue stars show results from unconstrained minimization; red boxes show results of the constrained minimization. 
Red and blue lines are guides to the eye. The IP is found as the negative of the HOMO energy. 
The black dotted line corresponds to the ideal correlation between an exact calculation and experiment.} \label{B3LYP_graph}
\end{center}
\end{figure}

The results for the calculated HOMO energies are plotted against the experimental results in Fig.~\ref{LDA_graph} for LDA,
Fig.~\ref{PBE_graph} for PBE and Fig.~\ref{B3LYP_graph} for B3LYP. 
From these plots, it is clear that the results of all the unconstrained methods give poorer fits to the experimental results than the constrained,
 with the latter being closer to the ideal correlation between calculation and experiment. 
For all three approximate functionals, 
the calculated IP almost always underestimates the experimental IP. This well-known underestimation of the 
IP~\cite{zhang2007comparison} continues to be present, but substantially reduced, in the constrained results, except in a handful of cases. 
 
\begin{figure*}[t]
\centerline{    \includegraphics[width=0.65\textwidth,angle=0]{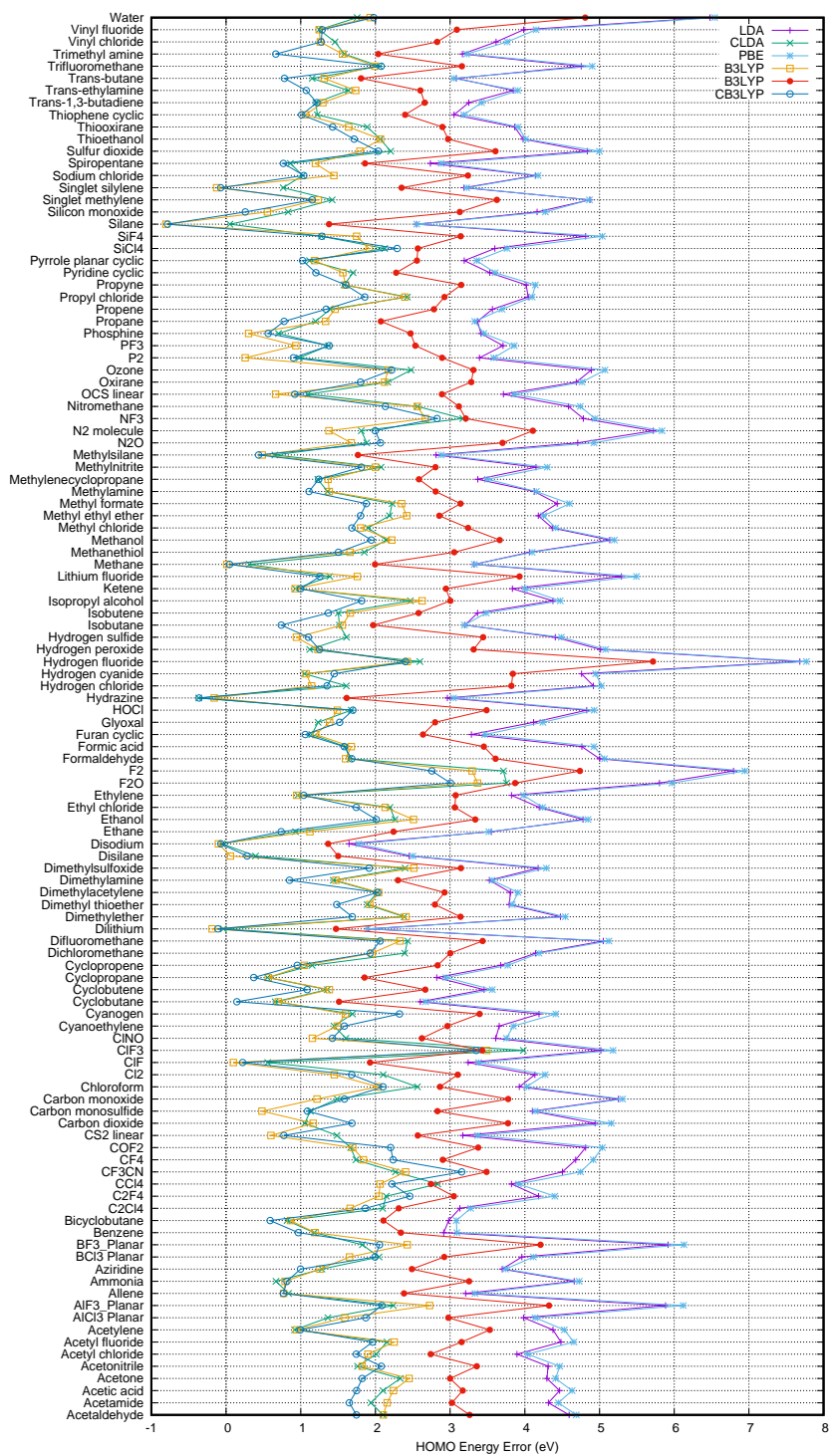} }
\caption{The differences between the calculated HOMO energy level and the experimental values for the ionization potential, 
comparing the unconstrained and constrained minimization of the LDA, PBE and B3LYP approximations. 
A positive value corresponds to an underestimation of the IP.} \label{HOMO_graph}
\end{figure*}

In Fig.~\ref{HOMO_graph}, we plot the error in the calculated ionization potential $\Delta \textrm{IP}$, where this error is given by 
the difference between the experimental and calculated values, $\Delta \textrm{IP} = 
\textrm{IP}_{\textrm{exp}} - \textrm{IP}_{\textrm{calc}}$.
A positive value in $\Delta \textrm{IP}$ implies an underestimation of the ionization potential. 
The inferior performance of the unconstrained relative to constrained minimization, is seen clearly in this figure, 
with IP errors of 4eV or more occurring frequently in the unconstrained case. 
The improvement of (unconstrained) B3LYP results over LDA and PBE is also evident, due to the partial cancellation of SIs in B3LYP. 
This improvement, however, is surpassed and offset by the constrained minimization technique to obtain the effective potential, 
with the three approximations giving similar results to each other. 


\begin{table}
\begin{center}
 \caption{The average error, $\bar{\Delta}$, standard deviation of the error, $\sigma$, average percentage error, $\bar{\delta}$, and standard deviation of the percentage error, $\bar{\sigma}$, from experimental results for the ionization potential (IP) for the molecules in Fig.~\ref{HOMO_graph}. The IP was approximated by the energy of the HOMO calculated using unconstrained functionals LDA, PBE, B3LYP and the constrained functionals CLDA, CPBE, CB3LYP. The average energy increase, $\Delta E$, of the total energies of the constrained calculations compared to the unconstrained are also shown.  }	\label{HOMO_table}		
\resizebox{\columnwidth}{!}{
\begin{tabular}{ccccccc} 
\noalign{\smallskip}\hline\noalign{\smallskip}
 & LDA & CLDA & PBE & CPBE & B3LYP & CB3LYP  \\
\noalign{\smallskip}\hline\noalign{\smallskip} 
$\bar{\Delta}$ (eV)  &	4.08	&	1.61	&	4.20	&	1.51	&	2.94	&	1.42	\\[2pt]
$\sigma$ (eV)  &	 0.93	&	0.74	&	0.94	&	0.77	&	0.71	&	0.73	\\[2pt]
$\bar{\delta}$  &	38\%&	15\%&	39\%&	14\%&	27\%&	13\%	\\[2pt] 
$\bar{\sigma}$  &	6\%&	6\%&	5\%&	7\%&	5\%&	6\%	\\[2pt]
$\Delta E$ (meV) &	& 0.1&&0.2&&0.3\\
\noalign{\smallskip}\hline\noalign{\smallskip}
 \end{tabular}
}
\end{center}
\end{table}	

A quantitative summary of the observations of the graphs in Figs.~\ref{LDA_graph} - 
\ref{HOMO_graph} can be found in Table \ref{HOMO_table}. There, we show
the average error, $\bar{\Delta}$, and the percentage error
$\bar{\delta}$, defined by averaging over the absolute value of 
$\Delta \textrm{IP}$ from Fig.~\ref{HOMO_graph}, and $|\Delta \textrm{IP}|/ \textrm{IP}$. The standard deviations $\sigma$ and $\bar{\sigma}$ of the absolute values of the $\Delta \textrm{IP}$ and $|\Delta \textrm{IP}|/ \textrm{IP}$ are also shown. 
The improvements of the constrained methods amount to a reduction in the average error for LDA and PBE by $\sim 2.5$eV while the 
B3LYP average error is halved to $\sim1.5$eV. 
For LDA and PBE these reductions correspond to a percentage improvement of 25\% and for B3LYP the improvement is 14\% over the 
unconstrained result. The standard deviation of the constrained results are smaller than the unconstrained for LDA and PBE or almost equal for B3LYP. The quality of the results improves not only because the average error decreases but also the standard deviation.

An important result that is evident in Fig.~\ref{HOMO_graph} and Table~\ref{HOMO_table} is the similarity of the results 
of the constrained optimizations, with all three approximations giving similar averages and similar deviations. 
One might expect this for the CLDA and CPBE calculations, since the unconstrained results are similar.
However, although the B3LYP results are shifted by approximately 1eV compared to the LDA and PBE results, the CB3LYP results show 
no such shift when compared to CLDA and CPBE. \\

As far as total energies are concerned, the replacement of the KS potential with the constrained effective one
is expected to raise the obtained total energies. In the last row of Table~\ref{HOMO_table} we show the 
 average increase in the total energy, $\Delta E$, from that of the corresponding KS calculation. We notice that the value of this increase is rather small. In other words, by enforcing 
the constraints of Eqs.~\eqref{constraint1}, \eqref{constraint2}, we obtain total energies very close to the unconstrained
KS minimum while on the other hand the orbital energies of the HOMO are substantially improved. As we have mentioned, the price is that the optimal potential is no longer the functional derivative of the potential energy with respect to the electron density. An interesting question of course is whether there exists a modified total energy functional that yields the obtained effective potential as its functional derivative w.r.t. the density.
The almost negligible size of the total energy raise for the constrained calculation
 is consistent with the observation\cite{grits} that 
potential terms with minimal influence in the total energy are responsible for the large deviation of the HOMO energies from the IPs.
Thus, a viable path for correcting the HOMO energies is the identification and correction for
such erroneous terms, as we aim to do in this work.

Another consequence of the constraint of Eq~\eqref{constraint1} is the introduction of a weak size inconsistency. Since 
the increase of the total energy for the constrained minimization has a very 
small value, the size inconsistency for the total energies is also minor. 
The effect on IPs on the other hand is more pronounced especially for small systems and goes to zero as the size of the constituent systems increases.

Calculations were also performed on a set of closed-shell anions where the IP coincides with the electron affinity (EA) of the neutral system.
The advantages over unconstrained functionals can be clearly seen, these results are found in Table~\ref{Anion_table}. 
Due to the expected diffuse nature of the HOMO in anions the augmented cc-pVTZ orbital basis set was used. 
With most approximate density functionals, the HOMO of the ions is found positive, i.e. they are predicted to have unbound electrons in most cases. 
This is a well known failure of many density functional approximations. 
With the constrained minimization method, we find that the same density functional approximations correctly predict that these anions have bound electrons, 
in agreement with experimental results. These results demonstrate that the improvements in the ionization energies are not limited to neutral molecules but 
can also be applied to anions.

\begin{table*}
\begin{center}
 \caption{The calculated IPs (in eV) for a set of anions using both constrained and unconstrained methods for the functionals LDA, PBE, B3LYP compared with experimental values for the electron affinities of the neutral systems. The average error, $\bar{\Delta}$, the average percentage error, $\bar{\delta}$, and the average increase in the total energies, $\Delta E$,  are shown for each of the functionals.}	\label{Anion_table}		
\begin{tabular}{cccccccc} 
\hline\noalign{\smallskip}
 system & LDA & CLDA & PBE & CPBE & B3LYP & CB3LYP  & Exp\\
\noalign{\smallskip}\hline\noalign{\smallskip} 
CH$_3^-$&-&0.30&-&0.26&-&0.51&0.08\\
CN$^-$&0.17&2.96&0.05&2.78&1.33&3.45&3.86\\
Cl$^-$&-&2.62&-&2.63&0.86&3.07&3.61\\
F$^-$&-&2.24&-&2.16&0.01&2.62&3.40\\
NH$_2^-$&-&0.23&-&0.15&-&0.50&0.77\\
OH$^-$&-&1.07&-&0.98&-&1.42&1.83\\
PH$_2^-$&-&0.74&-&0.75&-&0.91&1.27\\
SH$^-$&-&1.57&-&1.57&-&1.91&2.31\\
SiH$_3^-$&-&1.30&-&1.30&-&1.50&1.41\\
\noalign{\smallskip}\hline\noalign{\smallskip}
$\bar{\Delta}$ (eV) &&0.66&&0.70&&0.41&\\[2pt]
$\bar{\delta}\:(^*)$&	&35\%	&	&38\%	&	&20\% &\\[2pt]
$\Delta E$ (meV)& &0.015 & &0.052  & &0.15 &  \\
\noalign{\smallskip}\hline
\multicolumn{8}{c}{($^*$)The result for CH$_3^-$ is excluded as it dominates the percentage error.}
 \end{tabular}
\end{center}
\end{table*}	

\section{Conclusions}

We have presented in detail and investigated the performance of the method by Gidopoulos and Lathiotakis \cite{gidopoulos2012constraining} to remove SI effects 
from the effective KS potential, for three popular DFAs, LDA, PBE, and B3LYP.
A novelty of this method is the proposition that deficiencies of approximate KS potentials 
can be corrected by replacing the KS potentials with variationally optimized effective potentials that satisfy
certain properties. In our method, these properties are that the electron repulsion
density integrates to $N$-1 and is everywhere positive, Eqs.~\eqref{constraint1}, \eqref{constraint2}.  

The constrained minimization method was tested on its prediction for the ionization potential of  a large set of molecules. 
Based on  our results, the constrained method is found to offer substantial improvements for all approximate functionals tested, 
with a reduction of the average error for LDA from 4.08eV in the unconstrained case to 1.61eV with the constrained method. 
Similar reductions are found for PBE, while for the hybrid B3LYP functional the average error is almost halved from 2.94eV to 
1.42eV. We also applied the method to the calculation of the HOMO energies of a group of anions which were found correctly negative.
These energies, however, were found systematically smaller (by 20-38\%) than the electron affinities of the neutral system. 
In addition, we found that, in all cases, the imposition of the constraints only marginally affects the total energy of the system. Finally, we point out that the corrected IPs obtained with our method are still not very accurate, reflecting the limitations of the underlying DFAs. Improved results for the IPs can be obtained either by a more refined DFA or by directly modeling the effective single particle potential\cite{GLLB,SAOP,grits}.

These results show the importance of correcting for SI effects when calculating ionization potentials, and demonstrate 
the applicability of the constrained method in order to remove these self interaction effects in the KS potential. 
The constrained local potential is found to be a powerful method for improving the results of approximate 
functionals that contain self interactions. 

Importantly, the constrained minimization results appear to be independent of the particular approximation, 
as can be seen from Fig.~\ref{HOMO_graph} and Table~\ref{HOMO_table}, 
where the constrained optimization results for the three DFAs give similar results.
This property can be used to allow for more efficient calculations using a DFA that has a low computational cost but is 
of similar accuracy, once the constrained minimization method is used.

\section*{Acknowledgments}
The work was supported by The Leverhulme Trust, through a Research Project Grant with number RPG-2016-005. 
NNL acknowledges support by the project ``Advanced Materials and 
Devices'' (MIS 5002409) implemented under the ``Action for the Strategic 
Development on the Research and Technological Sector'', funded by the 
Operational Programme ``Competitiveness, Entrepreneurship and Innovation'' 
(NSRF 2014-2020) and co-financed by Greece and the European Union (European 
Regional Development Fund).

\bibliographystyle{apsrev}


\begin{thebibliography}{33}
\expandafter\ifx\csname natexlab\endcsname\relax\def\natexlab#1{#1}\fi
\expandafter\ifx\csname bibnamefont\endcsname\relax
  \def\bibnamefont#1{#1}\fi
\expandafter\ifx\csname bibfnamefont\endcsname\relax
  \def\bibfnamefont#1{#1}\fi
\expandafter\ifx\csname citenamefont\endcsname\relax
  \def\citenamefont#1{#1}\fi
\expandafter\ifx\csname url\endcsname\relax
  \def\url#1{\texttt{#1}}\fi
\expandafter\ifx\csname urlprefix\endcsname\relax\def\urlprefix{URL }\fi
\providecommand{\bibinfo}[2]{#2}
\providecommand{\eprint}[2][]{\url{#2}}

\bibitem[{\citenamefont{Perdew and Zunger}(1981)}]{pzsic}
\bibinfo{author}{\bibfnamefont{J.~P.} \bibnamefont{Perdew}} \bibnamefont{and}
  \bibinfo{author}{\bibfnamefont{A.}~\bibnamefont{Zunger}},
  \bibinfo{journal}{Physical Review B} \textbf{\bibinfo{volume}{23}},
  \bibinfo{pages}{5048} (\bibinfo{year}{1981}).

\bibitem[{\citenamefont{Blair et~al.}(2015)\citenamefont{Blair, Kroukis, and
  Gidopoulos}}]{hhfj}
\bibinfo{author}{\bibfnamefont{A.~I.} \bibnamefont{Blair}},
  \bibinfo{author}{\bibfnamefont{A.}~\bibnamefont{Kroukis}}, \bibnamefont{and}
  \bibinfo{author}{\bibfnamefont{I.~N.} \bibnamefont{Gidopoulos}},
  \bibinfo{journal}{J Chem Phys} \textbf{\bibinfo{volume}{142}},
  \bibinfo{pages}{084116} (\bibinfo{year}{2015}).

\bibitem[{\citenamefont{Lundberg and Siegbahn}(2005)}]{lundberg2005quantifying}
\bibinfo{author}{\bibfnamefont{M.}~\bibnamefont{Lundberg}} \bibnamefont{and}
  \bibinfo{author}{\bibfnamefont{P.~E.} \bibnamefont{Siegbahn}},
  \bibinfo{journal}{The Journal of Chemical Physics}
  \textbf{\bibinfo{volume}{122}}, \bibinfo{pages}{224103}
  (\bibinfo{year}{2005}).

\bibitem[{\citenamefont{R{\"o}sch and Trickey}(1997)}]{rosch1997comment}
\bibinfo{author}{\bibfnamefont{N.}~\bibnamefont{R{\"o}sch}} \bibnamefont{and}
  \bibinfo{author}{\bibfnamefont{S.}~\bibnamefont{Trickey}},
  \bibinfo{journal}{The Journal of Chemical Physics}
  \textbf{\bibinfo{volume}{106}}, \bibinfo{pages}{8940} (\bibinfo{year}{1997}).

\bibitem[{\citenamefont{Toher et~al.}(2005)\citenamefont{Toher, Filippetti,
  Sanvito, and Burke}}]{toher2005self}
\bibinfo{author}{\bibfnamefont{C.}~\bibnamefont{Toher}},
  \bibinfo{author}{\bibfnamefont{A.}~\bibnamefont{Filippetti}},
  \bibinfo{author}{\bibfnamefont{S.}~\bibnamefont{Sanvito}}, \bibnamefont{and}
  \bibinfo{author}{\bibfnamefont{K.}~\bibnamefont{Burke}},
  \bibinfo{journal}{Physical Review Letters} \textbf{\bibinfo{volume}{95}},
  \bibinfo{pages}{146402} (\bibinfo{year}{2005}).

\bibitem[{\citenamefont{Goedecker and Umrigar}(1997)}]{goedecker1997critical}
\bibinfo{author}{\bibfnamefont{S.}~\bibnamefont{Goedecker}} \bibnamefont{and}
  \bibinfo{author}{\bibfnamefont{C.}~\bibnamefont{Umrigar}},
  \bibinfo{journal}{Physical Review A} \textbf{\bibinfo{volume}{55}},
  \bibinfo{pages}{1765} (\bibinfo{year}{1997}).

\bibitem[{\citenamefont{Perdew and Levy}(1983)}]{perdew1983physical}
\bibinfo{author}{\bibfnamefont{J.~P.} \bibnamefont{Perdew}} \bibnamefont{and}
  \bibinfo{author}{\bibfnamefont{M.}~\bibnamefont{Levy}},
  \bibinfo{journal}{Physical Review Letters} \textbf{\bibinfo{volume}{51}},
  \bibinfo{pages}{1884} (\bibinfo{year}{1983}).

\bibitem[{\citenamefont{Almbladh and von Barth}(1985)}]{vonbarth}
\bibinfo{author}{\bibfnamefont{C.-O.} \bibnamefont{Almbladh}} \bibnamefont{and}
  \bibinfo{author}{\bibfnamefont{U.}~\bibnamefont{von Barth}},
  \bibinfo{journal}{Phys Rev B} \textbf{\bibinfo{volume}{31}},
  \bibinfo{pages}{3231} (\bibinfo{year}{1985}).

\bibitem[{\citenamefont{G\"orling}(1999)}]{Gor1999}
\bibinfo{author}{\bibfnamefont{A.}~\bibnamefont{G\"orling}},
  \bibinfo{journal}{Phys. Rev. Lett.} \textbf{\bibinfo{volume}{83}},
  \bibinfo{pages}{5459} (\bibinfo{year}{1999}).

\bibitem[{\citenamefont{Gidopoulos and
  Lathiotakis}(2012{\natexlab{a}})}]{gidopoulos2012constraining}
\bibinfo{author}{\bibfnamefont{N.~I.} \bibnamefont{Gidopoulos}}
  \bibnamefont{and} \bibinfo{author}{\bibfnamefont{N.~N.}
  \bibnamefont{Lathiotakis}}, \bibinfo{journal}{The Journal of chemical
  physics} \textbf{\bibinfo{volume}{136}}, \bibinfo{pages}{224109}
  (\bibinfo{year}{2012}{\natexlab{a}}).

\bibitem[{\citenamefont{Gidopoulos and
  Lathiotakis}(2015{\natexlab{a}})}]{GIDOPOULOS2015129}
\bibinfo{author}{\bibfnamefont{N.}~\bibnamefont{Gidopoulos}} \bibnamefont{and}
  \bibinfo{author}{\bibfnamefont{N.~N.} \bibnamefont{Lathiotakis}},
  \bibinfo{journal}{Advances In Atomic, Molecular, and Optical Physics}
  \textbf{\bibinfo{volume}{64}}, \bibinfo{pages}{129 }
  (\bibinfo{year}{2015}{\natexlab{a}}), ISSN \bibinfo{issn}{1049-250X}.

\bibitem[{\citenamefont{van Leeuwen and Baerends}(1994)}]{LB94}
\bibinfo{author}{\bibfnamefont{R.}~\bibnamefont{van Leeuwen}} \bibnamefont{and}
  \bibinfo{author}{\bibfnamefont{E.~J.} \bibnamefont{Baerends}},
  \bibinfo{journal}{Phys. Rev. A} \textbf{\bibinfo{volume}{49}},
  \bibinfo{pages}{2421} (\bibinfo{year}{1994}).

\bibitem[{\citenamefont{Legrand et~al.}(2002)\citenamefont{Legrand, Suraud, and
  Reinhard}}]{ADSIC}
\bibinfo{author}{\bibfnamefont{C.}~\bibnamefont{Legrand}},
  \bibinfo{author}{\bibfnamefont{E.}~\bibnamefont{Suraud}}, \bibnamefont{and}
  \bibinfo{author}{\bibfnamefont{P.-G.} \bibnamefont{Reinhard}},
  \bibinfo{journal}{Journal of Physics B: Atomic, Molecular and Optical
  Physics} \textbf{\bibinfo{volume}{35}}, \bibinfo{pages}{1115}
  (\bibinfo{year}{2002}).

\bibitem[{\citenamefont{Tsuneda and Hirao}(2014)}]{tsuneda2014self}
\bibinfo{author}{\bibfnamefont{T.}~\bibnamefont{Tsuneda}} \bibnamefont{and}
  \bibinfo{author}{\bibfnamefont{K.}~\bibnamefont{Hirao}},
  \bibinfo{journal}{The Journal of chemical physics}
  \textbf{\bibinfo{volume}{140}}, \bibinfo{pages}{18A513}
  (\bibinfo{year}{2014}).

\bibitem[{\citenamefont{Pederson et~al.}(2014)\citenamefont{Pederson,
  Ruzsinszky, and Perdew}}]{pederson2014communication}
\bibinfo{author}{\bibfnamefont{M.~R.} \bibnamefont{Pederson}},
  \bibinfo{author}{\bibfnamefont{A.}~\bibnamefont{Ruzsinszky}},
  \bibnamefont{and} \bibinfo{author}{\bibfnamefont{J.~P.}
  \bibnamefont{Perdew}}, \bibinfo{journal}{The Journal of Chemical Physics}
  \textbf{\bibinfo{volume}{140}}, \bibinfo{pages}{121103}
  (\bibinfo{year}{2014}).

\bibitem[{\citenamefont{Gidopoulos and
  Lathiotakis}(2015{\natexlab{b}})}]{clda_review}
\bibinfo{author}{\bibfnamefont{N.}~\bibnamefont{Gidopoulos}} \bibnamefont{and}
  \bibinfo{author}{\bibfnamefont{N.~N.} \bibnamefont{Lathiotakis}},
  \bibinfo{journal}{Advances In Atomic, Molecular, and Optical Physics}
  \textbf{\bibinfo{volume}{64}}, \bibinfo{pages}{129 }
  (\bibinfo{year}{2015}{\natexlab{b}}), ISSN \bibinfo{issn}{1049-250X}.

\bibitem[{\citenamefont{Clark et~al.}(2017)\citenamefont{Clark, Hollins,
  Refson, and Gidopoulos}}]{clark_gidopoulos_sic}
\bibinfo{author}{\bibfnamefont{S.~J.} \bibnamefont{Clark}},
  \bibinfo{author}{\bibfnamefont{T.~W.} \bibnamefont{Hollins}},
  \bibinfo{author}{\bibfnamefont{K.}~\bibnamefont{Refson}}, \bibnamefont{and}
  \bibinfo{author}{\bibfnamefont{N.~I.} \bibnamefont{Gidopoulos}},
  \bibinfo{journal}{J. Phys.: Condens. Matter} \textbf{\bibinfo{volume}{00}},
  \bibinfo{pages}{8pp} (\bibinfo{year}{2017}).

\bibitem[{\citenamefont{K\"ummel and Perdew}(2003)}]{kummel_perdew}
\bibinfo{author}{\bibfnamefont{S.}~\bibnamefont{K\"ummel}} \bibnamefont{and}
  \bibinfo{author}{\bibfnamefont{J.~P.} \bibnamefont{Perdew}},
  \bibinfo{journal}{Molecular Physics} \textbf{\bibinfo{volume}{101}},
  \bibinfo{pages}{1363} (\bibinfo{year}{2003}).

\bibitem[{\citenamefont{Lathiotakis
  et~al.}(2014{\natexlab{a}})\citenamefont{Lathiotakis, Helbig, Rubio, and
  Gidopoulos}}]{localrdmft}
\bibinfo{author}{\bibfnamefont{N.~N.} \bibnamefont{Lathiotakis}},
  \bibinfo{author}{\bibfnamefont{N.}~\bibnamefont{Helbig}},
  \bibinfo{author}{\bibfnamefont{A.}~\bibnamefont{Rubio}}, \bibnamefont{and}
  \bibinfo{author}{\bibfnamefont{N.~I.} \bibnamefont{Gidopoulos}},
  \bibinfo{journal}{Phys. Rev. A} \textbf{\bibinfo{volume}{90}},
  \bibinfo{pages}{032511} (\bibinfo{year}{2014}{\natexlab{a}}).

\bibitem[{\citenamefont{Lathiotakis
  et~al.}(2014{\natexlab{b}})\citenamefont{Lathiotakis, Helbig, Rubio, and
  Gidopoulos}}]{localrdmftappl}
\bibinfo{author}{\bibfnamefont{N.~N.} \bibnamefont{Lathiotakis}},
  \bibinfo{author}{\bibfnamefont{N.}~\bibnamefont{Helbig}},
  \bibinfo{author}{\bibfnamefont{A.}~\bibnamefont{Rubio}}, \bibnamefont{and}
  \bibinfo{author}{\bibfnamefont{N.~I.} \bibnamefont{Gidopoulos}},
  \bibinfo{journal}{The Journal of Chemical Physics}
  \textbf{\bibinfo{volume}{141}}, \bibinfo{pages}{164120}
  (\bibinfo{year}{2014}{\natexlab{b}}).

\bibitem[{\citenamefont{Theophilou et~al.}(2015)\citenamefont{Theophilou,
  Lathiotakis, Gidopoulos, Rubio, and Helbig}}]{lrdmftorbs}
\bibinfo{author}{\bibfnamefont{I.}~\bibnamefont{Theophilou}},
  \bibinfo{author}{\bibfnamefont{N.~N.} \bibnamefont{Lathiotakis}},
  \bibinfo{author}{\bibfnamefont{N.~I.} \bibnamefont{Gidopoulos}},
  \bibinfo{author}{\bibfnamefont{A.}~\bibnamefont{Rubio}}, \bibnamefont{and}
  \bibinfo{author}{\bibfnamefont{N.}~\bibnamefont{Helbig}},
  \bibinfo{journal}{The Journal of Chemical Physics}
  \textbf{\bibinfo{volume}{143}}, \bibinfo{pages}{054106}
  (\bibinfo{year}{2015}).

\bibitem[{\citenamefont{Perdew et~al.}(1996)\citenamefont{Perdew, Burke, and
  Ernzerhof}}]{perdew1996generalized}
\bibinfo{author}{\bibfnamefont{J.~P.} \bibnamefont{Perdew}},
  \bibinfo{author}{\bibfnamefont{K.}~\bibnamefont{Burke}}, \bibnamefont{and}
  \bibinfo{author}{\bibfnamefont{M.}~\bibnamefont{Ernzerhof}},
  \bibinfo{journal}{Physical Review Letters} \textbf{\bibinfo{volume}{77}},
  \bibinfo{pages}{3865} (\bibinfo{year}{1996}).

\bibitem[{\citenamefont{Becke}(1993)}]{becke1993density}
\bibinfo{author}{\bibfnamefont{A.~D.} \bibnamefont{Becke}},
  \bibinfo{journal}{The Journal of chemical physics}
  \textbf{\bibinfo{volume}{98}}, \bibinfo{pages}{5648} (\bibinfo{year}{1993}).

\bibitem[{\citenamefont{Lee et~al.}(1988)\citenamefont{Lee, Yang, and
  Parr}}]{lee1988development}
\bibinfo{author}{\bibfnamefont{C.}~\bibnamefont{Lee}},
  \bibinfo{author}{\bibfnamefont{W.}~\bibnamefont{Yang}}, \bibnamefont{and}
  \bibinfo{author}{\bibfnamefont{R.~G.} \bibnamefont{Parr}},
  \bibinfo{journal}{Physical Review B} \textbf{\bibinfo{volume}{37}},
  \bibinfo{pages}{785} (\bibinfo{year}{1988}).

\bibitem[{\citenamefont{Zhan et~al.}(2003)\citenamefont{Zhan, Nichols, and
  Dixon}}]{zhan2003ionization}
\bibinfo{author}{\bibfnamefont{C.-G.} \bibnamefont{Zhan}},
  \bibinfo{author}{\bibfnamefont{J.~A.} \bibnamefont{Nichols}},
  \bibnamefont{and} \bibinfo{author}{\bibfnamefont{D.~A.} \bibnamefont{Dixon}},
  \bibinfo{journal}{The Journal of Physical Chemistry A}
  \textbf{\bibinfo{volume}{107}}, \bibinfo{pages}{4184} (\bibinfo{year}{2003}).

\bibitem[{\citenamefont{Hirata et~al.}(2001)\citenamefont{Hirata, Ivanov,
  Grabowski, Bartlett, Burke, and Talman}}]{talman_bartlett}
\bibinfo{author}{\bibfnamefont{S.}~\bibnamefont{Hirata}},
  \bibinfo{author}{\bibfnamefont{S.}~\bibnamefont{Ivanov}},
  \bibinfo{author}{\bibfnamefont{I.}~\bibnamefont{Grabowski}},
  \bibinfo{author}{\bibfnamefont{R.~J.} \bibnamefont{Bartlett}},
  \bibinfo{author}{\bibfnamefont{K.}~\bibnamefont{Burke}}, \bibnamefont{and}
  \bibinfo{author}{\bibfnamefont{J.~D.} \bibnamefont{Talman}},
  \bibinfo{journal}{J Chem Phys} \textbf{\bibinfo{volume}{115}},
  \bibinfo{pages}{1635} (\bibinfo{year}{2001}).

\bibitem[{\citenamefont{Gidopoulos and
  Lathiotakis}(2012{\natexlab{b}})}]{gl_nonanalyticity}
\bibinfo{author}{\bibfnamefont{N.~I.} \bibnamefont{Gidopoulos}}
  \bibnamefont{and} \bibinfo{author}{\bibfnamefont{N.~N.}
  \bibnamefont{Lathiotakis}}, \bibinfo{journal}{Phys. Rev. A}
  \textbf{\bibinfo{volume}{85}}, \bibinfo{pages}{052508}
  (\bibinfo{year}{2012}{\natexlab{b}}).

\bibitem[{\citenamefont{Lathiotakis and
  Marques}(2008)}]{lathiotakis2008benchmark}
\bibinfo{author}{\bibfnamefont{N.}~\bibnamefont{Lathiotakis}} \bibnamefont{and}
  \bibinfo{author}{\bibfnamefont{M.~A.} \bibnamefont{Marques}},
  \bibinfo{journal}{The Journal of Chemical Physics}
  \textbf{\bibinfo{volume}{128}}, \bibinfo{pages}{184103}
  (\bibinfo{year}{2008}).

\bibitem[{\citenamefont{Johnson~III}(2011)}]{johnson2005nist}
\bibinfo{author}{\bibfnamefont{R.~D.} \bibnamefont{Johnson~III}}
  (\bibinfo{year}{2011}), \urlprefix\url{http://cccbdb.nist.gov}.

\bibitem[{\citenamefont{Zhang and Musgrave}(2007)}]{zhang2007comparison}
\bibinfo{author}{\bibfnamefont{G.}~\bibnamefont{Zhang}} \bibnamefont{and}
  \bibinfo{author}{\bibfnamefont{C.~B.} \bibnamefont{Musgrave}},
  \bibinfo{journal}{The Journal of Physical Chemistry A}
  \textbf{\bibinfo{volume}{111}}, \bibinfo{pages}{1554} (\bibinfo{year}{2007}).

\bibitem[{\citenamefont{Gritsenko et~al.}(2016)\citenamefont{Gritsenko, Mentel,
  and Baerends}}]{grits}
\bibinfo{author}{\bibfnamefont{O.~V.} \bibnamefont{Gritsenko}},
  \bibinfo{author}{\bibfnamefont{L.~M.} \bibnamefont{Mentel}},
  \bibnamefont{and} \bibinfo{author}{\bibfnamefont{E.~J.}
  \bibnamefont{Baerends}}, \bibinfo{journal}{The Journal of Chemical Physics}
  \textbf{\bibinfo{volume}{144}}, \bibinfo{pages}{204114}
  (\bibinfo{year}{2016}).

\bibitem[{\citenamefont{Gritsenko et~al.}(1995)\citenamefont{Gritsenko, van
  Leeuwen, van Lenthe, and Baerends}}]{GLLB}
\bibinfo{author}{\bibfnamefont{O.}~\bibnamefont{Gritsenko}},
  \bibinfo{author}{\bibfnamefont{R.}~\bibnamefont{van Leeuwen}},
  \bibinfo{author}{\bibfnamefont{E.}~\bibnamefont{van Lenthe}},
  \bibnamefont{and} \bibinfo{author}{\bibfnamefont{E.~J.}
  \bibnamefont{Baerends}}, \bibinfo{journal}{Phys. Rev. A}
  \textbf{\bibinfo{volume}{51}}, \bibinfo{pages}{1944} (\bibinfo{year}{1995}).

\bibitem[{\citenamefont{Schipper et~al.}(2000)\citenamefont{Schipper,
  Gritsenko, Van~Gisbergen, and Baerends}}]{SAOP}
\bibinfo{author}{\bibfnamefont{P.}~\bibnamefont{Schipper}},
  \bibinfo{author}{\bibfnamefont{O.}~\bibnamefont{Gritsenko}},
  \bibinfo{author}{\bibfnamefont{S.}~\bibnamefont{Van~Gisbergen}},
  \bibnamefont{and} \bibinfo{author}{\bibfnamefont{E.}~\bibnamefont{Baerends}},
  \bibinfo{journal}{The Journal of Chemical Physics}
  \textbf{\bibinfo{volume}{112}}, \bibinfo{pages}{1344} (\bibinfo{year}{2000}).

\end{thebibliography}

\end{document}